%% file: CO2_1.tex
\documentclass[
aip,
jcp,%
amsmath,amssymb,
reprint,%
]{revtex4-1}

\usepackage{hyperref}
\pdfoutput=1
\usepackage{graphicx}
\usepackage{amssymb}
\usepackage{amsmath}
\usepackage{blkarray}
\usepackage{multirow}
\usepackage{natbib}
\usepackage{epstopdf}
\usepackage{float}

\usepackage{graphicx}
\usepackage{dcolumn}
\usepackage{bm}

\usepackage{tabu} 

\usepackage{xr}
\usepackage{color}
\usepackage{soul}

\begin{document}


\title{CO$_2$ packing polymorphism under pressure: mechanism and thermodynamics of the I-III polymorphic transition}

\author{Ilaria Gimondi}
\affiliation{Thomas Young Centre and Department of Chemical Engineering, University College London, London WC1E 7JE, UK.}%
\author{Matteo Salvalaglio}%
\email{m.salvalaglio@ucl.ac.uk}
\affiliation{Thomas Young Centre and Department of Chemical Engineering, University College London, London WC1E 7JE, UK.}%

\date{\today}

\begin{abstract}
In this work we describe the thermodynamics and mechanism of CO$_2$ polymorphic transitions under pressure from form I to form III combining standard molecular dynamics, well-tempered metadynamics and committor analysis. We find that the phase transformation takes place through a concerted rearrangement of CO$_2$ molecules, which unfolds via an anisotropic expansion of the CO$_2$ supercell. Furthermore, at high pressures we find that defected form I configurations are thermodynamically more stable with respect to form I without structural defects. Our computational approach shows the capability of simultaneously providing an extensive sampling of the configurational space, estimates of the thermodynamic stability and a suitable description of a complex, collective polymorphic transition mechanism.
\end{abstract}

\maketitle

\section{\label{sec:intro}Introduction}
\input{polyNaim.tex}
\input{CO2phasediagram.tex}

\section{\label{sec:methods}Methods}
\input{methods_short.tex}

\subsection*{Force field}	 
\input{ff2.tex}
\subsection*{Simulation setup}
\input{setup.tex}
\subsection*{Committor Analysis} 
\input{methods_part2new.tex}
\subsection*{Collective variables} 
\input{CVs.tex}

\section{\label{sec:results}Results}

\subsection{Free Energy of the I-III polymorphic transition as a function of pressure}
\input{FES.tex}

\subsection{Committor analysis}
\input{MFEP.tex}
\input{committor_new.tex} 
\input{mechanism.tex}

\section{\label{sec:conslusions}Conclusions}
\input{conclusions.tex}

\section*{Supplementary Material}
See supplementary material for further details on the collective variables, and additional results on MFEP, committor analysis, and unbiased simulations. 

\section*{\label{sec:acknowledgments}Acknowledgements}
The        authors        acknowledge        EPSRC        (Engineering        and        Physical        Sciences        Research        Council)        for        PhD        scholarship,        and        UCL       
Legion        High        Performance        Computing        Facility                for        access        to        Legion@UCL        and        associated        support        services,        in        the       
completion       of       this       work.

\section*{\label{sec:references}References}
\bibliographystyle{unsrt}
\bibliography{CO2_1}

\end{document}



\title{Supporting Information: CO$_2$ packing polymorphism under pressure: mechanism and thermodynamics of the I-III polymorphic transition}
\thanks{Footnote to title of article.}

\author{Ilaria Gimondi}
\affiliation{ 
Department of Chemical Engineering, University College London.
}%
\author{Matteo Salvalaglio}%
 \email{m.salvalaglio@ucl.ac.uk}
\affiliation{ 
Department of Chemical Engineering, University College London.
}%

\date{\today}

\maketitle

\section*{\label{sec:SI}Supporting Info}
\input{SI_test.tex}

\section{\label{sec:references}References}
\bibliographystyle{unsrt}
\bibliography{CO2_1}

%% file: polyNaim.tex
Polymorphism, namely the possibility that molecular crystals assemble in the solid phase in different crystal lattices, is ubiquitous in nature. The spatial arrangement of molecules is key in defining mechanical, physical, chemical, and functional properties of materials.  Understanding the molecular details of the thermodynamics and mechanisms underlying polymorphism is therefore key to develop detailed, rational descriptions of many natural and industrial processes \cite{Cruz-Cabeza2015,Cruz-Cabeza2014,Price2008,Price2015,Zykova-Timan2008,Valdes-Aguilera1989,Vishweshwar2005,Bauer2001}. 

In this direction, a notable effort is put in developing both \textit{ab initio} and enhanced sampling techniques to predict polymorphs of a molecule (in particular, CSP techniques \cite{Price2008,Bond2016,Reilly2016,Dunitz2004}), to evaluate their relative stability at finite temperature and pressure, i.e. at conditions relevant for the life-cycle of a solid product, and to study transition mechanism and kinetics.
Among enhanced sampling techniques, metadynamics\cite{Laio2002,Martonak2003,Martonak2005,Martonak2007,Ceriani2004,Zipoli2004,Karamertzanis2008,Raiteri2005,Zykova-Timan2008,Lukinov2015} (MetaD) and adiabatic  free  energy  dynamics\cite{Yu2011,Yu2014,Schneider2016}  (AFED) are employed in literature to study polymorphism.  
Indeed, over the years these techniques have been tested, developed and compared on benchmark systems and combined with CSP methods. Such works made a successful step towards the characterisation of solid phase transition, proving these tools to be powerful in the prediction of new structures and transition pathways as well as of the phase diagram without any \textit{a priori} knowledge. \newline However, a complete and systematic investigation of polymorphic transitions is still challenging.

In this work our aim is to exploit state of the art enhanced sampling simulations to investigate the thermodynamics and transition mechanisms at play in polymorphic transitions. To this aim, we combine well-tempered MetaD and committor analysis in order to identify a suitable low dimensional description of the transition between two polymorphic phases in collective variable space. To do so, here we focus our attention on solid CO$_2$, more precisely on the $I$-$III$ polymorphic transition that characterises CO$_2$ \textit{packing polymorphism}. \textit{Packing polymorphism} arises when two solid phases differ in the packing of molecules, which have all the same molecular structure, as opposite to \textit{conformational polymorphism}\cite{Cruz-Cabeza2014}.  

In molecular solid phases, CO$_2$ molecules maintain their gas phase conformation. To a first, crude, approximation, each molecule can in fact be described as rigid and the bending of the O-C-O 180$^{\circ}$ angle can be reasonably neglected. Thanks to such limited conformational flexibility, it is as easy as spontaneous to identify each CO$_2$ molecule with a vector passing through its axis; moreover, the centre of mass corresponds to the carbon atom at every simulation time. As a result, the state of each molecule can be completely characterised by the position of its centre of mass and the vector representing its orientation in space. 

%% file: CO2phasediagram.tex
\begin{figure*}[t]
\includegraphics[width=1\textwidth]{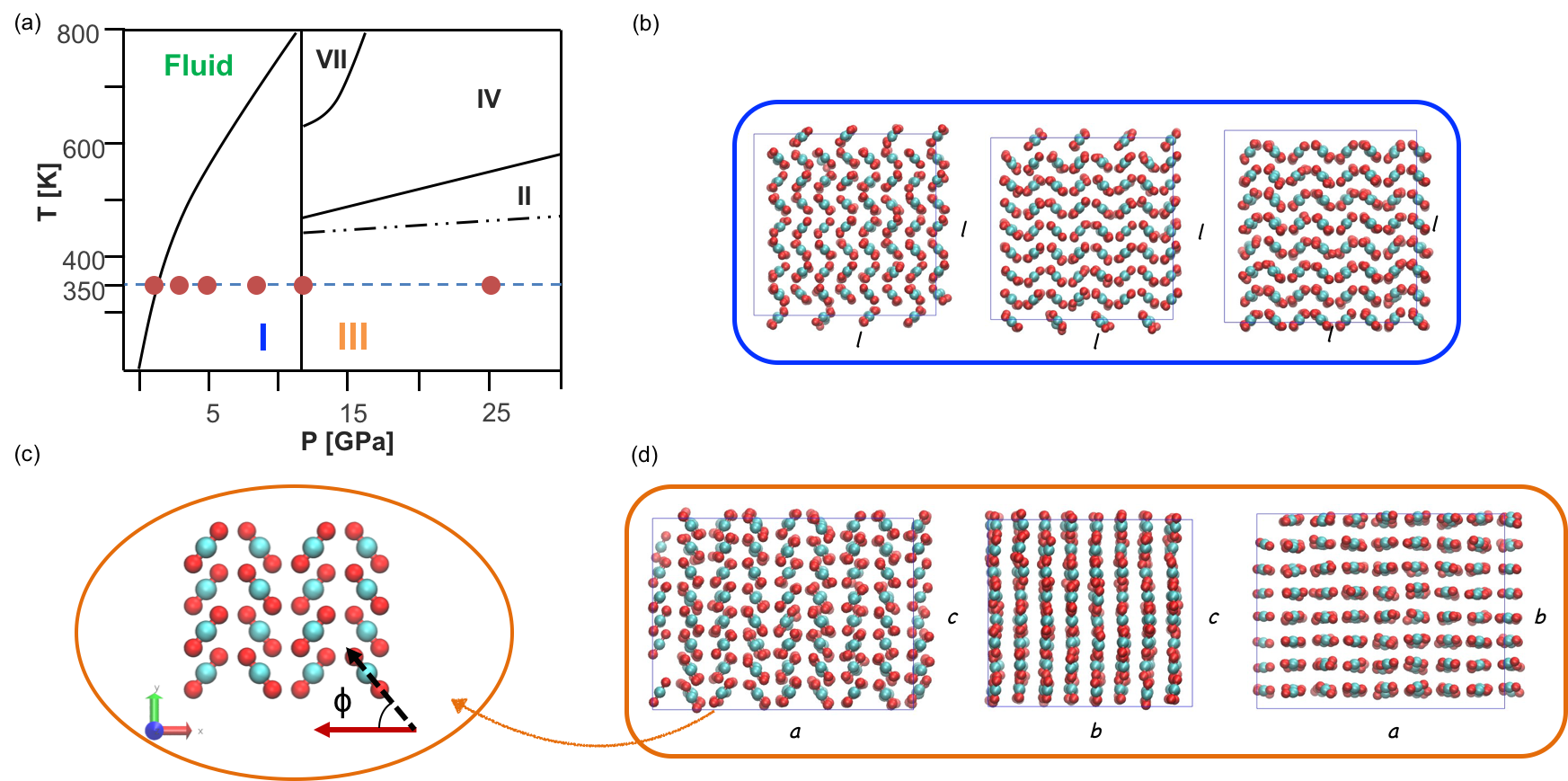}
\caption{(a) Detail of the phase diagram of CO$_2$ at high temperature and pressure from Datchi at al.\cite{Datchi2009} with highlighted the phases of interest of the present work (I in blue, III in orange). The red dots represent the condition of temperature and pressure investigated. (b) and (d) Snapshots of different planes of the 256-molecule super cell in phase I and III, respectively. (c) Detail of phase III with the typical 52-degree angle $\phi$ highlighted; in particular the red arrow aligns with the direction of the side of the box, while the black dashed one with the CO$_2$ molecular axis.}
\label{imageDiagr}
\end{figure*}

Despite its simple molecular structure, CO$_2$ has a rather complex solid-state phase diagram\cite{Datchi2009} (partially reported in Figure~\ref{imageDiagr} (a)). Indeed, at high temperature and pressure, seven different crystal structures have been detected so far, among which many are still debated\cite{Datchi2009,Datchi2012,Santoro2006,Yoo1999,Yoo2002,Park2003,Yoo2013,Giordano2007,Iota2001,Datchi2016,Iota2007,Shieh2013,Aoki1994,Bonev2003,Kuchta1988,Etters1989,Kuchta1993,Li2013,Hirata2014}.   The first form detected was molecular phase I, also called dry ice, crystallised directly from the melt; phase III followed, obtained through the compression of dry ice; the discovery of a polymeric structure, classified as phase V, attracted more interest to the study of this system, resulting in the identification of two more phases, II and IV. Phases II and IV are currently object of discussion as different groups hold contrasting views on their nature and role in the transition between molecular to non-molecular phases\cite{Yoo2002,Bonev2003,Datchi2016,Park2003,Yoo2013,Santoro2006,Gorelli2004,Datchi2009}.  
 Furthermore, an amorphous phase (VI) is also identified, and the existence of molecular form VII as a phase itself is still under investigation (see Figure~\ref{imageDiagr} (a)).  

In this work we study the transition between phases I and III, which are largely accepted and well characterised in the literature (Figure~\ref{imageDiagr} (b) and (d)). Their structural arrangements appear to hold several similarities. Both polymorphs, indeed, are face centred with four molecules in the unit cell, but while polymorph I's lattice is cubic Pa$\overline{3}$, III's is orthorombic \textit{Cmca}. A major difference is the orientation of the CO$_2$ particles: in phase I the molecular axis is in fact aligned with the diagonal of the cell, while in phase III they are arranged in parallel layers in which molecules describe a characteristic 52$^{\circ}$ angle, $\phi$, with the side of the lattice (Figure~\ref{imageDiagr} (c)).\\  
The $I-III$ transformation takes place at around 11-12 (11.8) GPa \cite{Datchi2009} independently of temperature ($\frac{dP_{I \leftrightarrow III}}{dT} = 0 $); nevertheless defining the transition conditions is a difficult task, and the pressure transition range is suggested to be wider (7 - 15 GPa)\cite{Shieh2013}, while Olijnyk et al.\cite{Olijnyk1988} observe transition III to I under unloading at around 2.5 - 4.5 GPa at 80 K. 
There is good agreement on the occurrence of a hysteresis of the specific volume, which decreases of about 2\%  from I to III \cite{Aoki1994,Bonev2003}. It is also generally accepted that the transition takes place through a concerted rotation of the molecules together with a deformation of the cubic structure to a parallelepiped, thanks to the peculiar geometrical features of the two phases. 
\\
The early works by Kuchta and Etters \cite{Kuchta1988,Etters1989,Kuchta1993} study such transition through NPT Monte Carlo (MC) simulations coupled with equalization of the Gibbs free energy in the phases under investigation, not without uncertainties\cite{Kuchta1993}. Moreover, the authors identify the orientation of the molecules in the lattice as the most relevant feature changing in the transition and thus they employ it as a transition coordinate to estimate the free energy profile associated to the transformation; their calculations both at 0 K\cite{Kuchta1988} and room temperature\cite{Kuchta1993} locate the transition pressure at 4.3 GPa. 
Li et al.\cite{Li2013} apply instead the second order M\o ller – Plesset (MP2) technique to the study of molecular crystals\cite{Hirata2014}. Despite small inaccuracies, they reach a good agreement about the transition pressure (around 11.9 - 12.7 GPa), obtained evaluating the free energy of the two polymorphs at different T-P conditions.

Here we aim at complementing the state of the art by providing an extensive sampling of the configurational space explored during the I - III polymorphic transition while contextually identifying the dominant transition mechanism. In the first part we perform well-tempered metadynamics simulations\cite{Barducci2008} with two order parameters as collective variables over a range of pressure at 350 K; from these, we obtain free energy surfaces that allow an insight into the relative stability between phase I and III under pressure. In the second part, we identify the most probable transition pathway and validate it through a committor analysis and a histogram test\cite{Peters2016,Tuckerman2010} on the transition state; we then propose a quantitative value for the energy barrier and a mechanism of the transition that takes into account quantitatively the reorientation of CO$_2$ molecules in the crystal as well as the deformation of the box. 

%% file: methods_short.tex
To circumvent timescale limitations of standard molecular dynamics, enhanced sampling techniques are designed to accelerate the sampling of rare events. In this work we employ well-tempered metadynamics\cite{Barducci2011} (WTMetaD). Briefly, WTMetaD is based on the introduction of a history-dependent bias potential (V$_G$) along a low-dimensional set of collective variables (CVs)\cite{Abrams2013,Barducci2011,Barducci2008,Laio2008,Valsson2016}.
Such bias allows for an efficient sampling of phase space, enhancing the escape from long-lived metastable states. Significantly, this result is achieved with little \textit{a priori} knowledge of the free energy landscape, and provides an estimate of the unbiased free energy surface (FES, F(\textbf{S})). 
For a detailed description of WTMetaD we refer the interested reader to Barducci et al.\cite{Barducci2008,Barducci2011}, and Valsson et al.\cite{Valsson2016}, and for a brief overview of its applications in crystallisation studies to Giberti et al.\cite{Giberti2015}.

%% file: ff2.tex
Here, we employ the rigid three-site TraPPE force field\cite{POTOFF1999,Potoff2001} (Table~\ref{table_TraPPE}), with Lennard Jones potential and Lorentz-Berthelot combination rules.
\newline This force field is chosen among a variety of models developed for CO$_2$\cite{Santoro2006,Datchi2009,Aimoli2014,Perez-Sanchez2013}, since, even if it is not tailor-made for the high temperature and high pressure regime of interest, it outperforms other models in the description not only of the liquid-vapor equilibrium at high pressures\cite{Aimoli2014} (up to 100 MPa), of the melting curve of dry ice (up to 1 GPa) and the triple point\cite{Perez-Sanchez2013}. Moreover, it has a better representation of the quadrupole, which is indeed relevant in carbon dioxide molecules and plays an important role in the solid phase stabilisation\cite{Price1987}.
\newline We employ two dummy atoms per molecule\cite{Sanghi2012} to mantain the desired rigidity and linearity of CO$_2$, avoiding instability caused by the rigid 180$^{\circ}$ OCO angle.  
The most relevant limitation of this model might be the rigidity of the CO$_2$ molecules\cite{Li2013}.

\begin{table}[h]
\centering
\setlength\extrarowheight{2pt}
\begin{tabular}{ ccccc }
 \hline
m$_C$ & m$_O$ & $\sigma_{C-C}$ [nm] & $\sigma_{O-O}$ [nm] & $\epsilon_{C-C}$ [kJ/mol] \\ 
 \hline
 12 & 16 & 0.280 & 0.305 & 0.224 \\[0.5ex] 
\hline
\hline
$\epsilon_{O-O}$ [kJ/mol] &  q$_c$ [e]  & q$_o$ [e] & $l_{C-O}$ [\AA] & $\alpha_{O-C-O}$ [$^{\circ}$]\\
\hline
0.657 &  0.70 & -0.35 & 1.160 & 180\\
\hline
\end{tabular}
\caption{Parameters for the TraPPE force field}
\label{table_TraPPE}
\end{table}

%% file: setup.tex
Long-range corrections for the Van der Waals interactions are included through the \textit{particle mesh Edwald} (\textit{pme}) method.
From consistency checks on the effect of the cut-off value on the system volume and energy when coupled with \textit{pme}, we find that 0.7 nm is a good trade-off between accuracy and computational cost. 

Isothermal and isobaric (NPT) simulations use Bussi-Donadio-Parrinello thermostat \cite{Bussi2007} and Berendsen anisotropic barostat\cite{Berendsen1984}  for T and P control, respectively. 
 The timestep employed is 0.5 fs.

For WTMetaD, the initial height of the Gaussians is 10 kJ/mol, with width 7.81e-3 for both CVs. The biasfactor is either 100 or 200  to allow the exploration of a wide portion of the phase-space. Moreover, we limit the elongation of each box side at 1.7 to 3.0 nm through the introduction of a repulsive potential.  
This action prevents an excessive and irreversible distortion of the box when the transition to melt is observed under anisotropic control. We highlight that such restraints are active only when the system undergoes large fluctuations in the liquid state.
The T-P conditions investigated include areas of the phase diagram where the most stable phase changes from melt to phase I to III: at 350 K, the range of pressure of the present study spans from 1 to 25 GPa (1, 3, 5, 8, 12, 25 GPa).
The initial configuration of WTMetaD simulations is phase I, initially equilibrated for 500 ps at NVT, then 5 ns NPT without \textit{pme} and additionally 5 ns NPT with \textit{pme}. All simulation boxes contain supercells of 256 CO$_2$ molecules. 
\newline We perform MD and WTMetaD simulations with Gromacs 5.2.1\cite{Abraham2015} and Plumed 2.2\cite{Tribello2014}; the building of the cells and the post processing of the data employs mainly VMD\cite{HUMP96}, to visualise trajectories, and MATLAB (R2015a). 

%% file: methods_part2new.tex
As mentioned in the opening, we complement our WTMetaD simulations with a committor analysis. While for a detailed description we refer to Tuckerman\cite{Tuckerman2010} and Peters\cite{Peters2016}, we recall here some useful definitions and procedures. 

The committor is defined as the probability $p_B(r_1,\ldots,r_N) \equiv p_B(\textbf{r})$ that a trajectory initiated from a configuration $r_1,\ldots , r_N \equiv \textbf{r}$ with velocities sampled from a Maxwell-Boltzmann distribution will arrive in state \textit{B} before state \textit{A}\cite{Tuckerman2010}. 
In our study we identify A as phase I and B as III. An important point on the pathway connecting two basins is the transition state (TS, indicated with *), which is the ensemble of configurations \textbf{r} with CV \textbf{S}(\textbf{r})=\textbf{S*} that have committor $p_B(\textbf{r})=0.5$; on free energy hypersurfaces, it corresponds to a saddle point, i.e. the highest energy state along the minimum energy path connecting two basins.  
\newline To locate the saddle point, we extract 135 configurations along the transition pathway and for each of them we run between 10 to 40 unbiased NPT simulations with different initial velocities randomly generated from a Maxwell-Boltzmann distribution. Simulations are  stopped when they commit either basin I or III and they are assigned an outcome value of 0 or 1, respectively. The average of the outcome values obtained from the set of trajectories generated for a given configuration provides an estimate of the committor $p_{III}(\textbf{r})$ for that configuration.

We have further analysed the \textit{histogram test}, which, instead, studies the committor distribution, $P(p_{B}(\textbf{r}))$, which is the probability that a configuration \textbf{r} with \textbf{S}(\textbf{r})=\textbf{S*} has committor $p_{B}(\textbf{r})=p^*$.
The shape of this distribution is a descriptor of the capability of the CVs to represent the transition mechanism: a Gaussian distribution results from good CVs, while a flat or parabolic distribution corresponds to CVs that do not describe adequately the transition state ensemble. 
\newline To evaluate the committor probability, we consider 41 configurations with CVs close to the estimated transition state, and for each of them, we evaluate the committor, $p_{III}^m$, as previously described, and build the histogram of $P(p_{III})$.

%% file: CVs.tex
\begin{figure*}[t]
\includegraphics[width=1\textwidth]{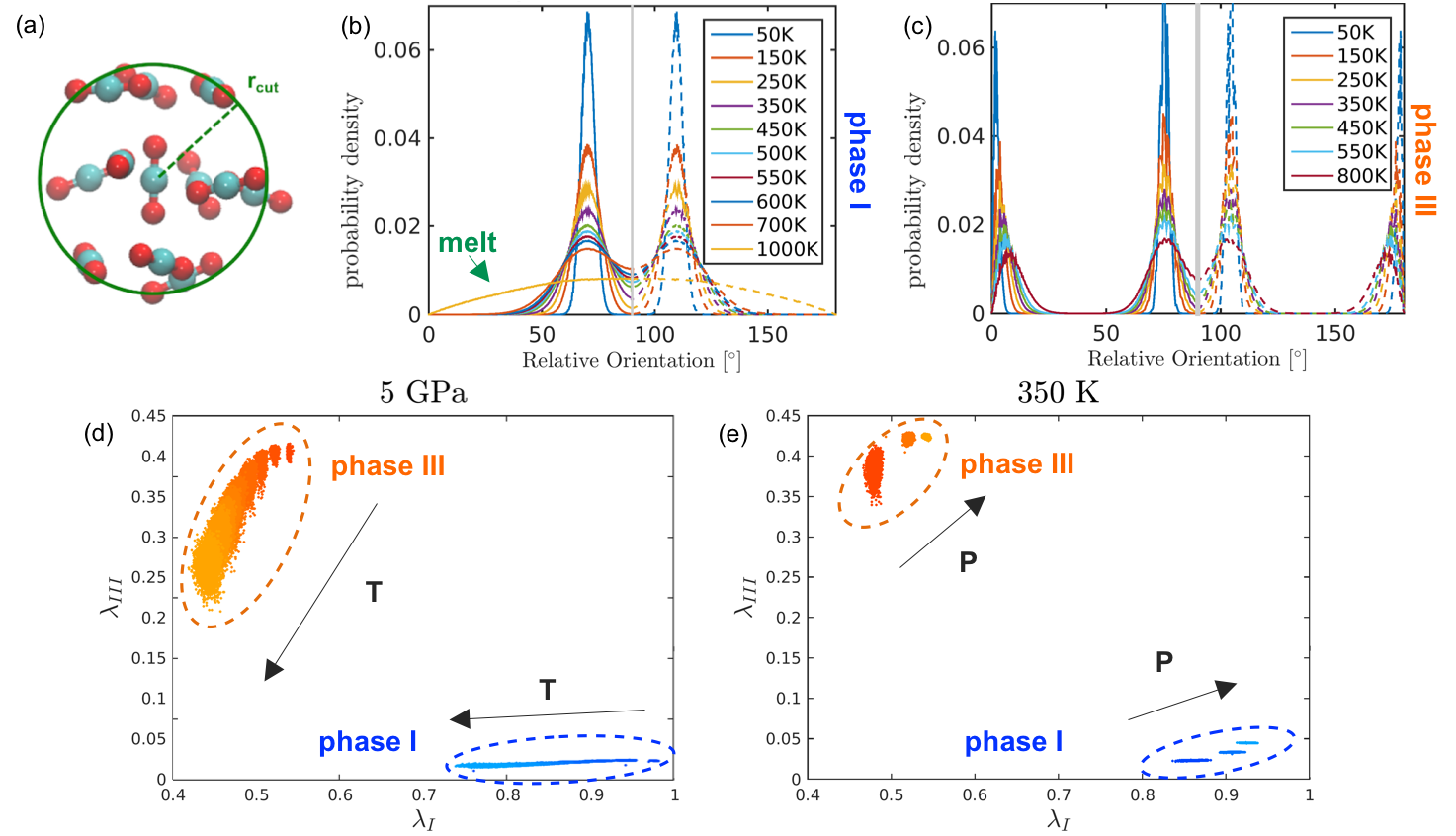}
\caption{(a) Visual model of the environment around a CO$_2$ molecule within the sphere of radius r$_{cut}$. (b) - (c) Angle distribution over a range of temperature for phases I at 5 GPa (b) and III at 25 GPa (c). The green arrow points an example for melt. (d) - (e) CV-space ($\lambda_I$-$\lambda_{III}$) for phase I, within the dotted blue line, and phase III, within the dotted orange line at 5 GPa  over a range of T from 50 to 1000 K (d) and at 350 K over a range of P from 5 to 25 GPa (e); the arrows point the direction of growing temperature or pressure.}
\label{imageCVs} 
\end{figure*}

In this work, we use a CV developed by Salvalaglio et al.\cite{Salvalaglio2012,Salvalaglio2013,Salvalaglio2015} and employed also in Giberti et al.\cite{Giberti2015_2}. In particular, every crystal structure has a unique typical local environment around each CO$_2$ molecule, a fingerprint of the arrangment, and this order parameter, hereafter called $\lambda$, exploits this feature to effectively distinguish polymorphs. Indeed, $\lambda$ describes crystallinity, a global property of the ensemble, as the sum of local contributions, $\Gamma_i$; each $\Gamma_i$ takes into account both the local density, $\rho_i$, within a cut-off, $r_{cut}$, around the $i$-th molecule, and the orientation, $\theta_{ij}$, respect to its neighbours (Figure~\ref{imageCVs} (a)). The value of $\lambda$ ranges between 0 and 1, as it expresses the portion of molecules in the system that are ordered according the geometry of a defined polymorphic structure.
\\
A complete description of the formulation of this parameter is reported in the $Supporting$ $Information$ and the cited literature.

From the characterization of the local order in polymorphs I and III we can observe and compare peculiarities of the angle distribution of each phase (Figure~\ref{imageCVs} (b) and (c)), useful in the following tuning of CVs. First of all, the arrangements of phases I and III present similarities, as there is overlap between the distributions of two characteristic angles, which are however centred in different values (in around 70.2$^\circ$ and 109.8$^\circ$ for form I, while in around  75.6$^\circ$ and 104.4$^\circ$ for III). Moreover, phase III populates two additional characteristic angles, with values smaller than 10$^\circ$ and bigger than 170$^\circ$, which might relate to the presence of layers. We remark also that the melt has a sinusoidal distribution of angles, consistent with a random orientation of molecules. As a final note, increasing temperature enhances the fluctuations of the molecules in the crystal without modifying the mean value of the characteristic angles; an exception to this are the layer angles of form III that, instead, change from $\sim$1$^\circ$ to $\sim$8$^\circ$ and from $\sim$179$^\circ$ to $\sim$172$^\circ$ with growing temperature.
\newline The number of neighbours in the first coordination shell shows, instead, a narrow distribution and the same value for the two structures, i.e. 12. Such observations lead to the tuning two CVs, namely $\lambda_I$ and $\lambda_{III}$ (see $Supporting$ $Information$).

\paragraph*{Order parameter $\lambda_I$} This CV expresses the degree of phase I-likeness. The purpose of the tuning is to maximize $\lambda_I$ when the crystal structure is phase I. To reach this aim, two characteristic angles, $\theta_{k}$, are included, which are the ones of phase I (Table~\ref{table_lambda}).

\paragraph*{Order parameter $\lambda_{III}$}
Similarly, the tuning of $\lambda_{III}$ aims at maximise the parameter in presence of phase III. However, in this case we do not talk about phase III-likeness, because as $\theta_{k}$ we select only the specific angles that characterize layers (Table~\ref{table_lambda}). 
 
For both CVs, the cut-off $r_{cut}$ is set to 4.0 \AA, as it delimits the first coordination shell; the width of the Gaussians associated to the angles, $\sigma_k$, instead, is in both cases set to maximise the difference between the value of $\lambda$ in phase I and melt and it is the same for both the characteristic angles due to the symmetry (Table~\ref{table_lambda}).
\begin{table}[h!]
\centering
\setlength\extrarowheight{5pt}
\begin{tabular}{ c|ccccc } 
  & $\theta_{1}$ [$^{\circ}$] & $\theta_{2}$ [$^{\circ}$] & $\sigma_{1}=\sigma_{2}$ [$^{\circ}$] & n$_{cut}$ [-] & r$_{cut}$ [\AA] \\
  \hline
   \hline
   $\lambda_I$ & 70.47 & 108.86 & 14.32 & 5 & 4 \\
$\lambda_{III}$ & 8.02 & 171.89 & 11.46 & 5 & 4 \\
\hline 
\end{tabular}
\caption{Tuning of the $\lambda$-order parameters.  The table reports both the angles set, $\theta_{1}$ and $\theta_{2}$, while only one Gaussian width, as for symmetry reasons it is the same for both angles.  The cut-off values for the number of neighbours, n$_{cut}$, and the coordination shell, r$_{cut}$, are presented as well.}
\label{table_lambda}
\end{table}

The phase-space evaluated on unbiased MD simulations suggests that the CVs are effective in the distinction of separate and well-defined areas for each phase (Figure~\ref{imageCVs} (d) and (e)). Furthermore, the average order parameters can be extracted as an ensemble average.

Temperature and pressure act on the location where phases are projected in CV-space: on the one hand, increasing temperature decreases the values of both $\lambda$s while widening their fluctuations, consistently with the fact that the volume increases and the molecules vibrate more; on the other hand, increasing pressure leads to an increase in the absolute value of the parameters, while narrowing their distribution.

%% file: FES.tex
In the following, we present the results of our study of the I - III polymorphic transition in CO$_2$.
\newline Firstly, we just mention that from preliminary MD unbiased simulations phase III has a smaller volume than phase I under all conditions investigated ($\sim$2\%, in agreement with experimental results); the volume predictions, however, slightly overestimate the experimental values (Figure~\ref{imageComp} in the \textit{Supporting Information}). In addition, for the same T-P settings, form I presents a lower potential energy than III: the potential energy of the system is thus not a good indicator of the relative thermodynamic stability at finite temperature. We report the outcome of MD in the \textit{Supporting Information}.

\begin{figure*}[t]
\includegraphics[width=1\textwidth]{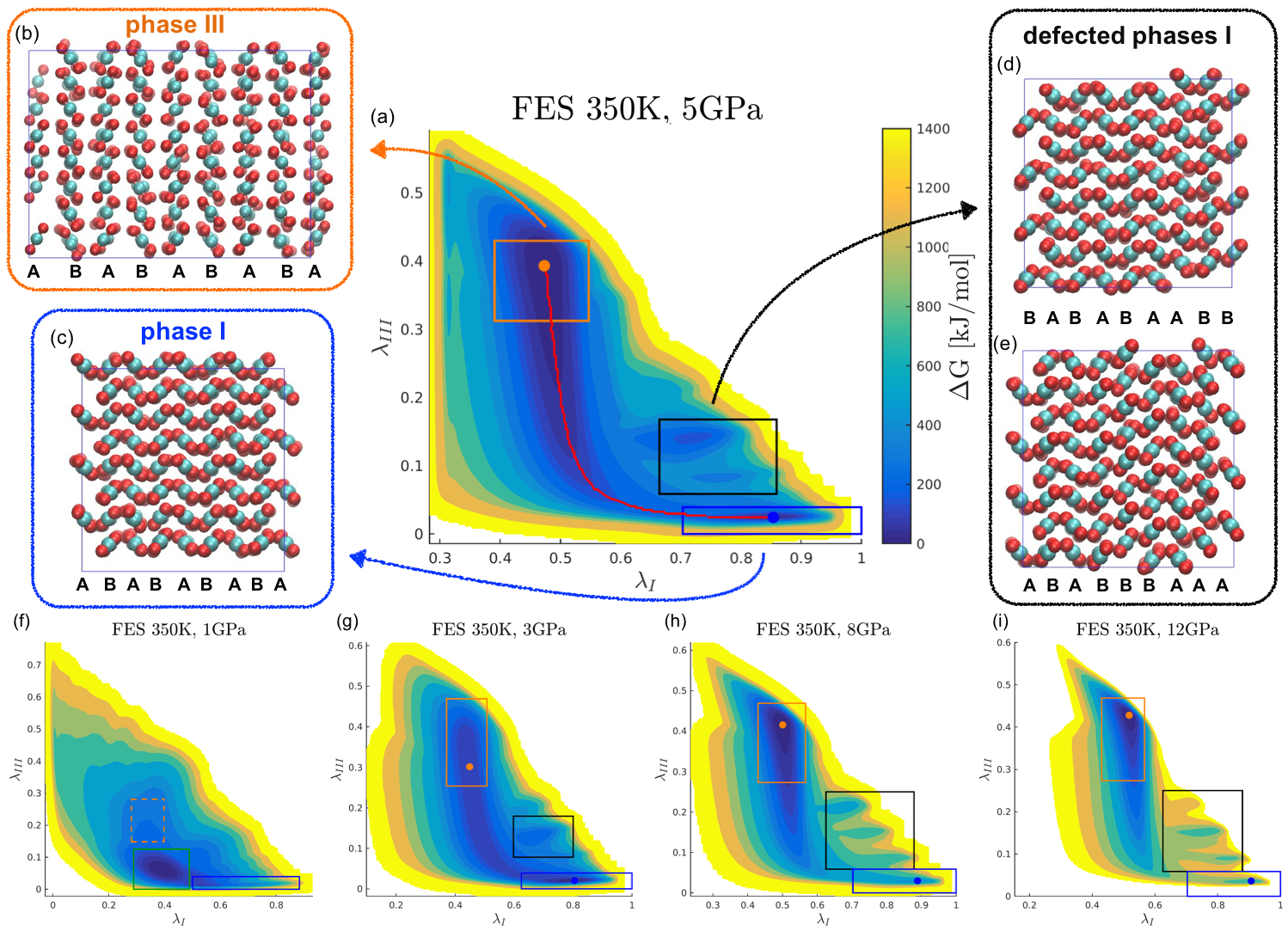}
\caption{FESs at 350 K under a range of pressure: 1 GPa (f), 3 GPa (g), 5 GPa (a), 8 GPa (h), and 12 GPa (i). The colour bar spaces for all surfaces from 0 to 1400 kJ/mol. Blue boxes locate the basin of phase I, orange phase III, melt is within a green box, while black rectangles identify phase I with defects. In addition, (a) reports also the minimum free energy transition path (red). The structures reported are III in (b), I in (c) and two examples of packing faults  in (d) and (e); the letters are an aid to compare the packing. Only one plane is displayed as the most explicative of defects; however, while one of the not shown planes is almost perfect, the other has the correct motif, but the layers are not perfectly aligned.}
\label{imageFES}
\end{figure*}

Then, we discuss the results obtained from WTMetaD simulations run with the set-up discussed before.
\begin{figure}[t]
\includegraphics[width=0.5\textwidth]{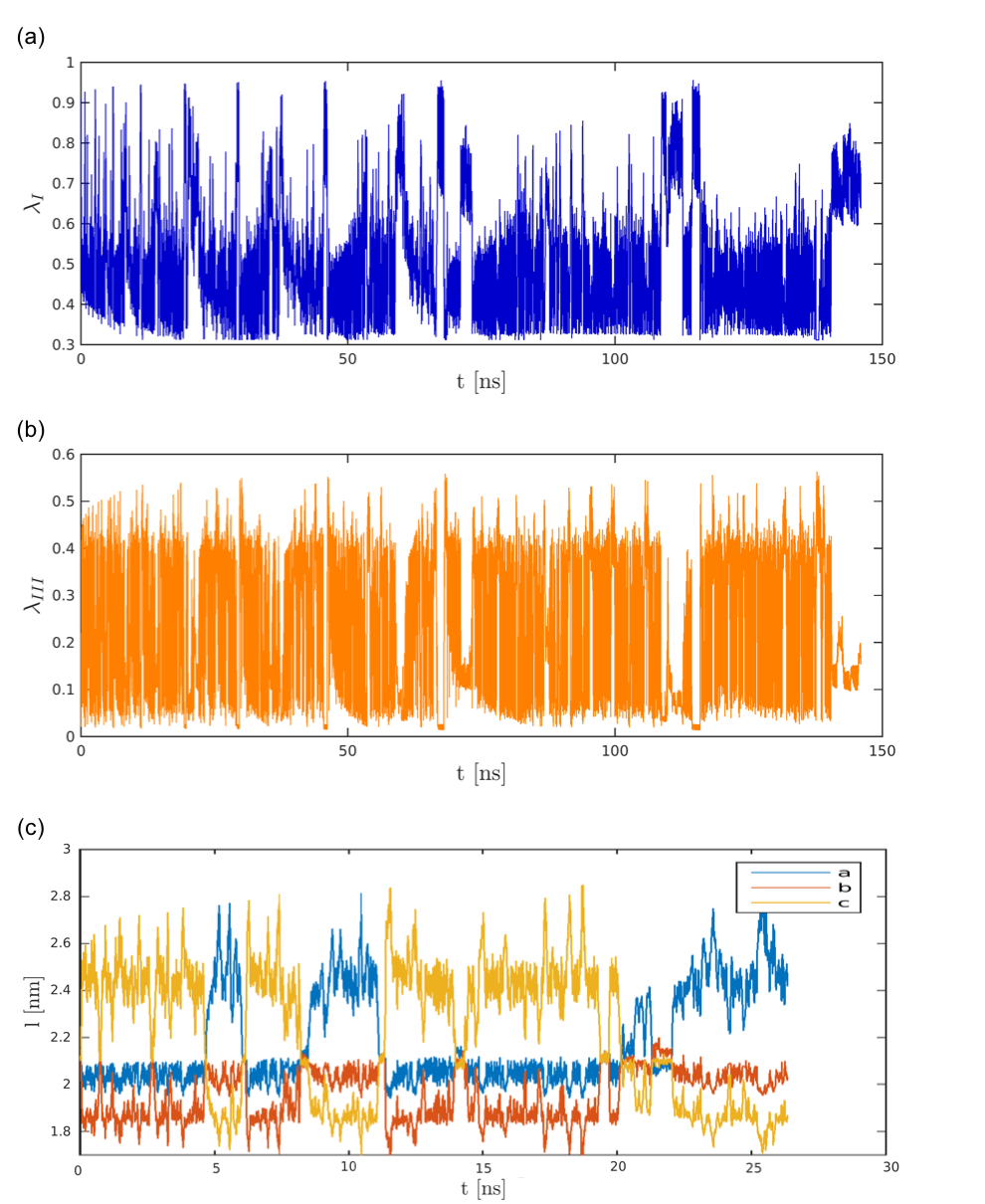}
\caption{Time evolution of the CVs $\lambda_I$ (a) and $\lambda_{III}$ (b) over 150 ns, and of the box edges (c) for the first 30 ns, of WTMetaD at 350 K - 5 GPa.}
\label{imagetevolution}
\end{figure}
\newline To begin with, we observe the temporal evolution of the CVs, for the explicative case at 350 K - 5 GPa (Figure~\ref{imagetevolution} (a) and (b)); the other conditions investigated (Figure~\ref{imageCons} in the \textit{Supporting Information}) behave in a reasonably similar way. In the plots in Figure~\ref{imagetevolution} it is possible to identify the system arranged in phase I as $\lambda_I$ (a) is high (fluctuations between 0.7 and 0.9), $\lambda_{III}$ (b) is below 0.05 and does not present relevant fluctuations, while the box edges (c) have the same length. The exploration of phase III's basin, instead, shows wider fluctuations in the range of 0.36 - 0.6 for $\lambda_I$ (a), and  0.1 - 0.4 for $\lambda_{III}$ (b); the box edges, moreover, fluctuates around the unbiased average. Thanks to this clarification, it is possible to spot in Figure~\ref{imagetevolution} that the system undergoes a significant number of recrossings between polymorphs I and III, in particular, four in slightly more than 5 ns at the beginning of the run. 
 In addition, the system explores areas of the CV-space which do not represent any of these polymorphs, feature that will result more evident from the plots of the free energy surface (Figure~\ref{imageFES}). On the same surfaces it will be possible to notice the important role that the mentioned fluctuations of the CVs have on the shape of the basins for the two phases. 

Furthermore, by observing the output trajectories and data of WTMetaD, we remark two interesting behaviours: on the one hand, CO$_2$ molecules in the simulation box rearrange with a concerted motion during a phase transition; on the other hand, we find that such transition is anisotropic, meaning that each side of the box is equally likely to either elongate or shorten from I to III (Figure~\ref{imagetevolution} (c)). In particular, this latter observation is important since by biasing as CVs order parameters that account for the spacial orientation of molecules, we obtain the consequential deformation of the supercell, without considering the box volume or edges as CVs, as instead done in previous MetaD works on polymorphism\cite{Zykova-Timan2008,Martonak2003,Martonak2005,Martonak2007,Ceriani2004,Karamertzanis2008,Raiteri2005,Lukinov2015}.

Next, we present the free energy surfaces (FES) reconstructed by WTMetaD. In such FESs, the free energy is expressed as a function of the CVs: G($\lambda_I$, $\lambda_{III}$) (Figure~\ref{imageFES} (a) and (f) to (i)). Before proceeding with the discussion, we underline that the FES at 25 GPa is not reported, as no recrossing is sampled from phase III; in addition, the results at 1 GPa are taken into account only qualitatively, since under such conditions phase III is so unstable that spontaneously evolves to I  in standard MD and it is thus not possible to locate its basin.
\newline Some considerations can be drawn from the study of the FESs. First of all, the location of the minima on the FES for phases I and III is accurately close to the prediction in Figure~\ref{imageCVs}(d)-(e). Moreover, phase III has a much wider basin than phase I and it develops mainly along $\lambda_{III}$, while phase I's mainly along $\lambda_I$, as underlined for the temporal evolution of the CVs (Figure~\ref{imagetevolution}). As mentioned before, the system explores a wide area of CV-space and, in particular, the presence of black boxes in Figure~\ref{imageFES} highlights the presence of defected phase I structures, which we shall analyse in detail later on. Relevant structural arrangements are reported in Figure~\ref{imageFES} (b) to (e).

\begin{figure*}[t]
\includegraphics[width=1\textwidth]{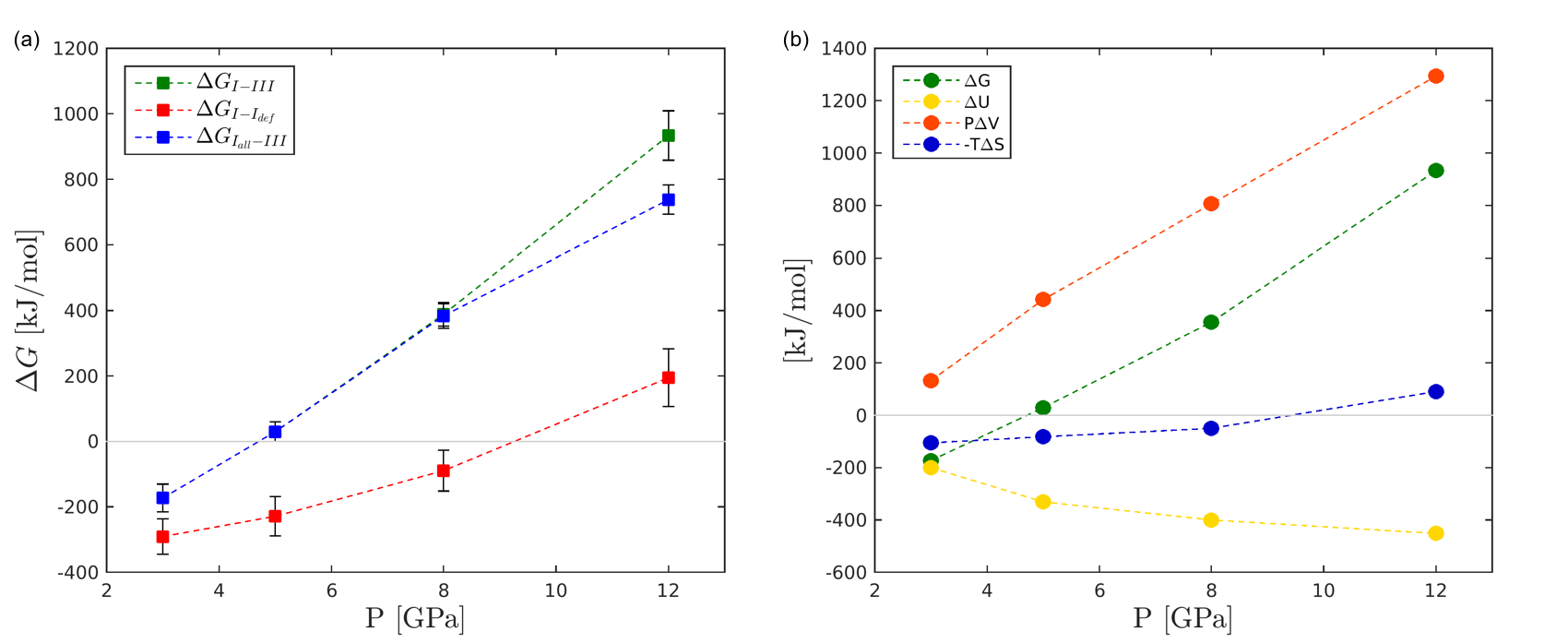}
\caption{Relative stability between different phases (a) and breakdown of the free energy (b) at 350 K and increasing pressure from 3 to 12 GPa. In (a) green squares represent $\Delta{G}$ between phase I and III, red between perfect phase I and defected structures I, while blue bewteen a comprehensive phase I, including configurations with and without defects,  and phase III. Positive values of $\Delta{G}$ mean that phase III (green, blue) and defected I (red) are more stable. The error bars are obtained from a weighted averaged on simulation time, similarly to Berteotti et al.\cite{Berteotti2011}. In (b) the focus in on the contributions to the relative stability between phases I and III: the $\Delta{G}$ obtained by WTMetaD is again plotted in green, yellow represents the internal energy difference of the two phases from MD, red their difference in mechanical work from MD, while blue the entropy difference obtained from the definition of Gibbs free energy (Eq~\ref{Gibbs}); we report the entropic contribution as -T$\Delta{S}$, so that for all the terms considered negative values stabilize phase I and positive phase III. In both graphs, dashed lines are an aid to the eye to visualise the trend.}
\label{imagestability} 
\end{figure*}

In order to compare the results of WTMetaD with the experimental phase diagram, we study quantitatively the relative stability between polymorphs.
\newline Keeping in mind that the free energy is a function of the probability distribution of the CVs, it is possible to evaluate $\Delta{G}_{I-III}$ as \eqref{deltaG}:
\begin{align}
\Delta{G}_{I-III} &=G_I-G_{III}=-\beta^{-1}\ln \left( \frac{p_{pI}}{p_{pIII}}\right) 
\label{deltaG}
\end{align}
Where $p_{pI}$ is the probability of phase I, $p_{pIII}$ of phase III, and $\beta$ is 1/kT.
The probability of each phase is computed as the integral of the distribution within the basin it occupies on the CV-space:
\begin{equation}
p_{pI} = \int_I p(\mathbf{\lambda})\,d\mathbf{\lambda} = \iint_{\lambda_I,\lambda_{III}\in I}p(\lambda_I,\lambda_{III})\,d\lambda_I\,d\lambda_{III}
\label{p_I}
\end{equation}
\begin{equation}
p_{pIII} = \int_{III} p(\mathbf{\lambda})\,d\mathbf{\lambda} = \iint_{\lambda_I,\lambda_{III}\in III}p(\lambda_I,\lambda_{III})\,d\lambda_I\,d\lambda_{III}
\label{p_III}
\end{equation}
The integration domains are identified by coloured boxes on the FES in Figure~\ref{imageFES} (a) and (f) to (i).
 In Figure~\ref{imagestability} (a) we report relevant $\Delta{G}$ values over the range of pressure considered. 
The relative stabilities in Figure~\ref{imagestability} (a) together with the FESs in Figure~\ref{imageFES} allow to draw some considerations on the phase diagram.
We observe that while the boundary of the solid - melt transition is in good agreement with experiments, the I - III transition pressure appears underestimated.  
From the $\Delta{G}_{I-III}$ pattern shown in green in Figure~\ref{imagestability} (a), the transition pressure at 350 K can be estimated as around 4.5 GPa. Despite underestimating the experimental value, the transition pressure agrees with literature results obtained treating CO$_2$ as a rigid molecule\cite{Kuchta1988,Etters1989,Kuchta1993}. We also recall that commonly experimental works rather than a single value report a transition pressure interval (see Introduction), to which our estimation is closer.
\newline Nevertheless, it is possible to notice that WTMetaD simulations are able to represent the overall trend observed in the phase diagram: increasing pressures increase the stability of phase III, while at decreasing values of pressure phase I is more stable, ultimately reaching the boundary with melt.
\newline Since the behaviour of solid carbon dioxide is so well described, it is possible to consider a translation of the phase diagram.

A further step in the analysis of the I - III relative stability is the breakdown of the free energy in its internal energy, mechanical work and entropy contributions. With this aim, we firstly evaluate the difference in internal energy, $\Delta{U}$ and mechanical work, P$\Delta{V}$, between phase I and phase III from the ensemble averages computed from the unbiased MD simulations; the entropy is thus obtained from the macroscopic definition of Gibbs free energy:
\begin{equation}
\Delta{G}=\Delta{U}+P\Delta{V}-T\Delta{S}
\label{Gibbs}
\end{equation}
From the results in Figure~\ref{imagestability} (b) it possible to notice some major features. First of all, the internal energy, $\Delta{U}$, stabilizes form I, while P$\Delta{V}$ is significant in the stabilisation of phase III; in both cases their contribution becomes more relevant with growing pressures.  The entropic term, instead, tends to favour form I, a part from pressure of 12 GPa.

\subsubsection*{Defected phases}
As mentioned in the analysis of the CVs (Figure~\ref{imagetevolution}) and of the FESs (Figure~\ref{imageFES}), at pressure equal and above 3 GPa, the system evolves to new, not \textit{a priori} known phases, which we recognise being defected structures I (Figure~\ref{imageFES} (d) and (e)). Indeed, such phases are  similar to phase I, being almost cubic and having  comparable arrangement; however, they display packing faults, \textit{planar defects} that break the orientation motif recognisable in perfect phase I. The perfect arrangement, in fact, presents the repetition of rows of CO$_2$ that alternate the orientation respect to the Cartesian axis in a sort of ABABABAB sequence (Figure~\ref{imageFES} (c)), while the defected phases replicate two or more lines with the same ``character'', AA or BB (Figure~\ref{imageFES} (d) and (e)). It is particularly remarkable the capability of WTMetaD to predict the production of defected structures, as, despite this phenomenon can takes place in experiments, it is ``underrepresented in the current literature''\cite{Sosso2016}, due to its difficult characterization both experimentally and through modelling.
\newline In addition, we observe that the stability of the defected phases increases at higher pressures, becoming ultimately even more stable than phase I ($\Delta{G}$ in red in Figure~\ref{imagestability} (a)). 
This behaviour may be due to a more difficult expansion of volume from phase III to I under higher pressure, and thus defected phases with a smaller volume form.

To complete the analysis of these phases, we run unbiased simulations under the same T-P conditions as the related WTMetaD, and with the defected arrangement of interest as initial configuration. The results show that these forms do not spontaneously undergo any transition: the creation and correction of defects is thus an activated event.

%% file: MFEP.tex
In the first part of this work, we have shown that our $\lambda$-order parameters are effective CVs in sampling the transition between polymorphs I and III, evaluating their relative stability, and exploring the phase space. In the following, we focus instead on the mechanism of the title transition. In particular, such analysis allows to evaluate the goodness of the CVs in the representation of the process, and to estimate quantitatively the transition pathway and the energy barrier to overcome. 

First of all, we characterize the minimum free energy path (MFEP) that connects the free energy minima corresponding to phase I and III. 
 The MFEP provides a representation in CV-space of the most probable set of intermediate states involved in the transition. Furthermore, the free energy profile along this path yields an estimate of the free energy barrier associated to the polymorphic transition. 
  As initial estimate of the MFEP we propose an approximation obtained as the combination of the projection of FES along the CVs, more precisely, of basin I along $\lambda_I$ and of basin III along $\lambda_{III}$, due to the observe typical L-shaped FES; further details about these approximations are provided in the \textit{Supporting Information}. We then employ such path as educated guess for an optimisation routine that enables to obtain the actual MFEP from a series of trial moves, whose acceptance is based on the free energy value. The algorithm is robust and the path converges to the same route from different and less educated initial guesses. 
\begin{figure*}[t]
\includegraphics[width=1\textwidth]{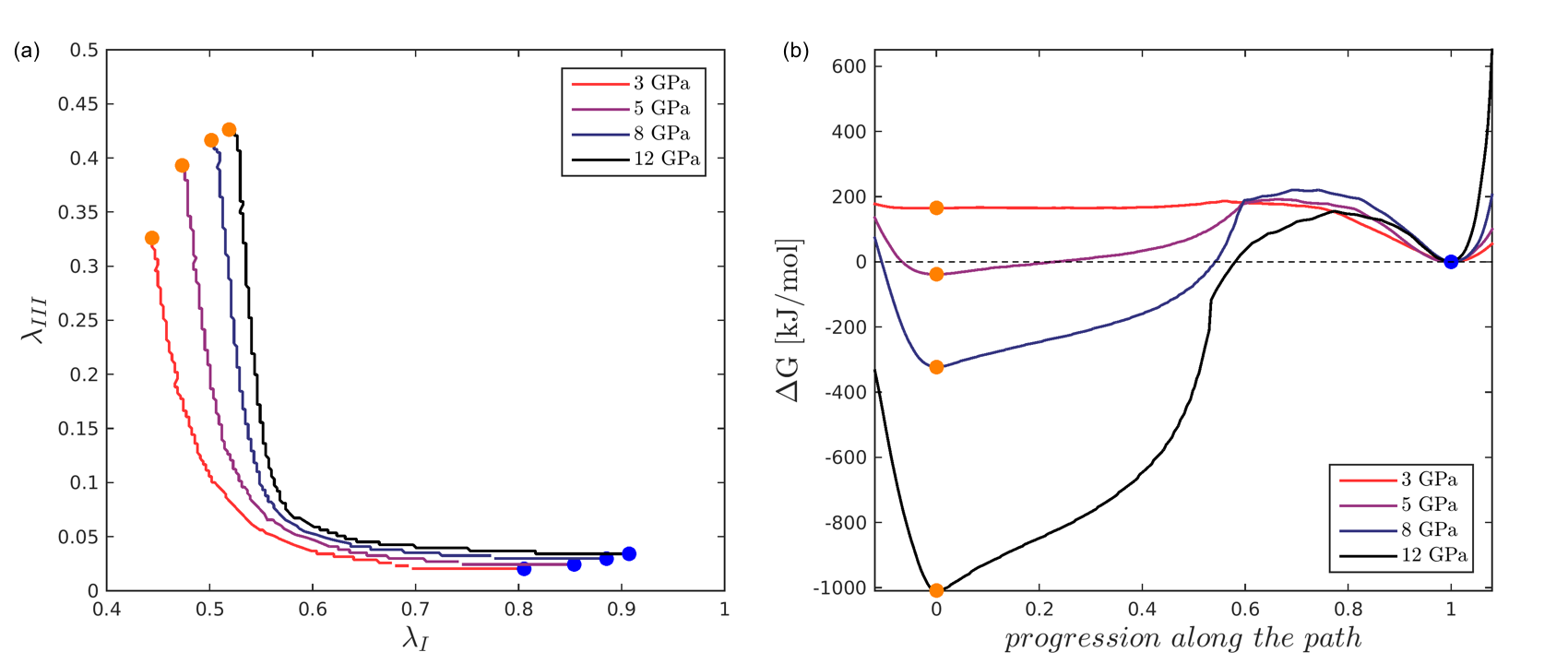}
\caption{(a) Transition pathway in the space of CVs at 350 K over the range of pressures investigated. (b) Projection of the free energy along the curvilinear path coordinate at 350 K over a range of pressures. The progression along the MFEP spaces between 0 in phase III and 1 in phase I. The minimum of phase I's basin is the free energy reference. In both graphs blue dots represent phase I, while  orange III.}
\label{imagepath}
\end{figure*}

In Figure~\ref{imageFES} (a) we report the MFEP on the FES at 350 K - 5 GPa, while in Figure~\ref{imagepath}(a) we compare transition pathways evaluated at different pressures. Interestingly, pressure only slightly affects the typical L-shape of the transition pathway, with the major difference being the location of the minima. Moreover, at this level of detail, the energy barrier to overcome from phase I to III appears similar at all pressures investigated (Figure~\ref{imagepath}(b)). We also highlight that the MFEP converges much earlier than the simulation, and no alternative routes connecting polymorphs I - III arise (Figure S7 in the \textit{Supporting Information}).

%% file: committor_new.tex
\begin{figure*}[t]
\includegraphics[width=0.82\textwidth]{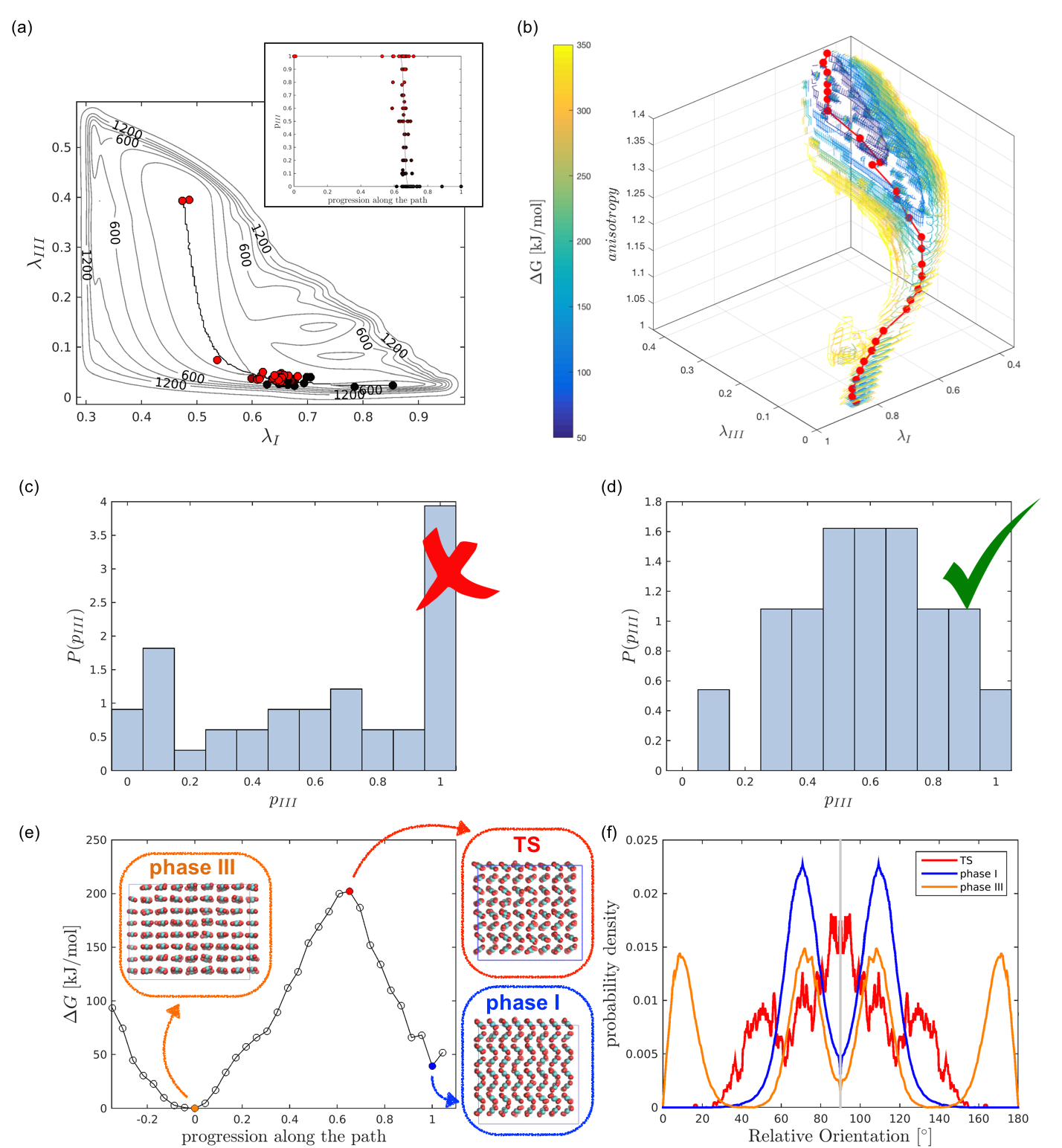} 
\caption{Committor analysis results, with the transition mechanism described only by the CVs (a,c), or by the CVs and the box anisotropy (b,d-f). In (a) coloured dots represent configurations, \textbf{r}, extracted along the MFEP (black) on the CV-space; their colour is based on the committor of each configuration, shading from black for $p_{III}(\textbf{r})$ = 0, to red for $p_{III}(\textbf{r})$ = 1. The same colour code applies to the insert in (a), which reports the committor as a function of the progression along the MFEP; in the same graph, the sigmoid dashed line is an aid to the eye to read the trend. The histogram test on configurations \textbf{r} so that \textbf{$\lambda$}(\textbf{r})=\textbf{$\lambda^*$} of the TS shows three peaks (c). The inclusion of the box anisotropy to the $\lambda$-order parameters to describe the mechanism requires a 3D representation and thus in (b) we report the free energy as a function of these three parameters through a colour plot and the transition path through red dots; the results of the histogram test (d) on configurations with the same $\lambda_I$, $\lambda_{III}$, and anisotropy of the TS confirms that this set of parameters is effective and complete. In (e) the free energy is plotted as a function of the progression along the 3D path highlighted in (b), with reference in phase III. An orange dot locates phase III (at progression zero along the path), a blue dot phase I (at progression 1), and a red dot the transition state; representations of the structures of phase I, phase III and the TS are within blue, orange and red rectangles, respectively. The same colour code is employed for the angle distributions for phase I, phase III and the TS presented in (f).}
\label{image_histo}
\end{figure*}

The next step towards a quantitative characterization of the transition mechanism is the validation of MFEP through a committor analysis\cite{Peters2016,Tuckerman2010}, with histogram test on the \emph{apparent} transition state. We discuss hereafter the explicative case of 350 K - 5 GPa.

To begin with, we locate the transition state by evaluating the committor of 135 configurations extracted along the MFEP. Interestingly, configurations with committor different from zero and one are not evenly distributed along the transition pathway, but grouped in a narrow area around the saddle point, estimated in $\lambda_I^*$ = 0.65, $\lambda_{III}^*$ = 0.034; as a result, the cumulative distribution along the path resembles a very steep $sigma$ shape (Figure~\ref{image_histo}(a)). The behaviour shown in Figure~\ref{image_histo}(a) suggests that the order parameters alone might not be enough to account for the transition mechanism. The validation proceeds with a histogram test on the saddle point. We evaluate the committor of 41 configurations with \textbf{$\lambda$}(\textbf{r}) around \textbf{$\lambda^*$} and represent the results on a histogram (Figure~\ref{image_histo} (c)). Such histogram shows three peaks, sign that our CVs alone are not effective reaction coordinates and other parameters need to be included in the mechanism description.

In order to identify the additional parameters to take into account, we deepen our analysis and further investigate the dependence of $p_{III}(\textbf{r})$ on properties such as potential energy, volume and box dimensions (Figure S6 in the \textit{Supporting Information}). The results suggest that the deformation of the lattice plays a role in the representation of I - III transition mechanism. 
We define this deformation through the simulation box \textit{anisotropy}, i.e. the ratio between the longest and the shortest sides of the cell; its value spans from 1 in cubic phase I to 1.35 in orthorhombic phase III. As a result, we note that only configurations \textbf{r} along the pathway with anisotropy of the box between 1.14 and 1.145 have committor non identical to 0 or 1, and, in particular the TS is uniquely located in $\lambda_I^*$ = 0.65, $\lambda_{III}^*$ = 0.034, $anisotropy^*$ = 1.1421, whise characteristic orientations are presented in Figure~\ref{image_histo} (f). We thus repeat the histogram test on 19 configurations with CVs and \textit{anisotropy} close to the TS and the outcome shows, as expected, a Gaussian shape (Figure~\ref{image_histo} (d)): as a result, to effectively describe the mechanism of the I - III transition of solid CO$_2$ all three parameters, namely $\lambda_I$, $\lambda_{III}$ and $anisotropy$, have to be taken into account.

We thus evaluate\cite{Bonomi2009} the free energy as a function of the three parameters of interest: $G(\lambda_I,\lambda_{III},anisotropy)$. On such FES we identify the 3D MFEP that connects phase I to phase III (Figure~\ref{image_histo} (b)): its projection on the $\lambda_I-\lambda_{III}$ plane reasonably overlaps with the MFEP previously evaluated; moreover the anisotropy of the box monotonically increases from I to III, and vice versa.

%% file: mechanism.tex
Summing up the analysis carried on in this second part of the work, the transition from cubic phase I to orthorhombic phase III can be thus described as the sequence of the following actions:
\begin{itemize}
\item  The CO$_2$ molecules firstly tend to distort the typical phase I lattice and, as a consequence, the value of $\lambda_I$ decreases, with no relevant increase of $\lambda_{III}$ (horizontal branch of the L-shaped pathway); at the same time the box starts deforming, elongating one side and reducing the others, thus increasing its anisotropy.
\item  Then, when the deformation of the cell reaches the anysotropy threshold value of the transition state, the system completes the  rearrangement to phase III; indeed, the molecules start organizing into parallel layers and the volume decreases. From this point the transformation proceeds on the vertical branch of the L-shaped pathway, with increasing $\lambda_{III}$ for relative small variations of $\lambda_I$.
\end{itemize}
The motion of the molecules in the crystal during the transition is concerted.

From the investigation of the 3D MFEP we obtain also quantitative information about the height of the barrier for the polymorphic transformation: for a system composed by 256 CO$_2$ molecules at 350 K - 5 GPa the transition state is located at about 202 kJ/mol (Figure~\ref{image_histo}(e)), with reference zero in phase III, i.e. the absolute minimum of the FES. 

%% file: conclusions.tex
In this work we presented an investigation of the I - III polymorphic transition in carbon dioxide under pressure. Our approach combines molecular dynamics, well-tempered metadynamics and committor analysis to provide a broad insight into this phenomenon.

Firstly, we performed WTMetaD simulations at 350 K over a range of pressure (1 - 25 GPa) with two order parameters as CVs. These parameters, $\lambda_I$ and $\lambda_{III}$, are built on the local order around each CO$_2$ molecule and account for the reorientation of the molecules in the crystal. This feature allows to clearly distinguish in the $\lambda_I$-$\lambda_{III}$ CV space configurations that belong to phase I or phase III and to  clearly resolve amorphous configurations. Moreover, metadynamics exploration with these CVs allows to sample the formation of packing faults in phase I. Interestingly, we observe the deformation of the cell, a \emph{global} rearrangement of the configuration, taking place as a consequence of enhanced sampling along  $\lambda_I$ and $\lambda_{III}$, which account for local order. This also permitted to notice that, in a I to III transition, all sides of the box have the same probability to elongate or shorten. 

From the FESs resulting from WTMetaD, we evaluated the free energy difference between polymorphs; we observed that the predicted trend of the I - III relative stability over pressure is in agreement with the carbon dioxide phase diagram: increasing pressures move from the melt-phase I boundary, to the region of stability of phase I, to the one phase III. 
We estimate the transition pressure at $\sim$4.5 GPa, in agreement with previous literature works that considered carbon dioxide as a rigid molecule\cite{Kuchta1988,Etters1989,Kuchta1993}. Furthermore, our model suggest that the stability of the defected configurations increases with pressure. While at low pressure undefected form I is more stable than the ensemble of its defected counterparts, at high pressure the latter appears to dominate.  

Alongside a description the I-III transition thermodynamics, we assess the I-III polymorphic transition mechanism for the representative case at 350 K - 5 GPa. To this aim we identify the MFEP connecting phase I to phase III in CV space and we validated the pathway carrying out committor analysis and histogram test on an ensemble of configurations corresponding to the saddle point in CV space. From this analysis it emerges that to quantitatively identify the transition mechanism we need to consider the anisotropic deformation of the $CO_2$ supercell alongside order parameters accounting for the local arrangement of $CO_2$ molecules. This analysis allowed to identify a reliable approximation of the transition pathway, and hence to quantify the free energy barrier associated with the transition.

Our work shows that, by combining opportunely designed order parameters with state-of-the-art enhanced sampling methods and committor analysis, we can provide an in-depth characterisation of both thermodynamics and transition mechanisms of polymorphic transformations at finite temperature.

%% file: SI_test.tex
\subsubsection{Collective Variables}
In this work, we run WTMetaD employing as CVs a set of two $\lambda$ order parameters, namely $\lambda_I$ and $\lambda_{III}$. Such parameters were first developed and used by Salvalaglio et al.\cite{Salvalaglio2012,Salvalaglio2013,Salvalaglio2015} and Giberti et al.\cite{Giberti2015_2} in the study of nucleation and melting of urea. We refer to these works while presenting here further details on the formulation of the parameters.

As mentioned in the main manuscript, $\lambda$ represents the degree of crystallinity of the system as a sum of local contributions, $\Gamma_i$ (Eq \eqref{lambda}). 
\begin{align}
\lambda = \frac{1}{N}\sum_{i=1}^{N}\Gamma_i
\label{lambda}
\end{align}
Dividing by the total number of molecules in the system, N, ensures that $\lambda$, similarly to a molecular fraction, expresses the portion of particles organized alike a defined crystal structure, ranging from 0 to 1.
\newline As said, each $\Gamma_i$ considers the local order around the \textit{i}-th CO$_2$ molecule in terms of density, $\rho_i$, and orientation with respect to neighbours, $\theta_{ij}$. First of all, the local density, $\rho_i$, is based on the coordination number, n$_i$: if the number of neighbours of \textit{i} is bigger than the \textit{cut-off} n$_{cut}$, the molecule is crystal like (Eq \eqref{rho}).
\begin{align}
\rho_i = \frac{1}{1+e^{-b \left( n_i - n_{cut} \right)}}
\label{rho}
\end{align}
where $b$ tunes the slope of the switching function and n$_i$ is a function of the distance $r_{ij}$ between the i-$th$ and the j-$th$ molecules and a well-defined cut-off $r_{cut}$ (Eq \eqref{ni} and Eq \eqref{fij}):
\begin{align}
n_i = \sum_{\substack{j=1  \\ j\neq i}}^{N}f_{ij}
\label{ni}
\end{align}
\begin{align}
f_{ij} = \frac{1}{1+e^{a \left(  r_{cut} - r_i \right)}}
\label{fij}
\end{align}
where a tunes the slope.
\newline A second feature considered is the orientation between neighbouring molecules, i.e. the angle between  \textit{i} and the \textit{j}-th molecules within $r_{cut}$ from \textit{i}, $\theta_{ij}$. This angle $\theta_{ij}$ in a crystal fluctuates around the characteristic orientations $\theta_{k}$ according to a Gaussian distribution (with width $\sigma_k$); the term $\Theta_{ij}$ accounts for this behaviour (Eq \eqref{Thij}). To consider only orientation between neighbours, $\Theta_{ij}$ is multiplied by $f_{ij}$ (Eq \eqref{fij}).
\begin{align}
\Theta_{ij} = \sum_{k = 1}^{k_{max}}e^{ - \frac{\left( \theta_{ij} - \theta_{k}   \right)^2}{2\sigma_k^2}    }
\label{Thij}
\end{align}
\newline Overall, the local crystallinity $\Gamma_i$ is expressed by Eq \eqref{Gammai} and tends to 1 when the molecule is in the descripted solid phase:
\begin{align}
\Gamma_{i} = \frac{\rho_i}{n_i}\sum_{j = 1}^{N}f_{ij}\Theta_{ij}
\label{Gammai}
\end{align}
The use of tunable switching functions allows to have continuous and differentiable CVs.

\begin{figure}[t]
\includegraphics[width=0.5\textwidth]{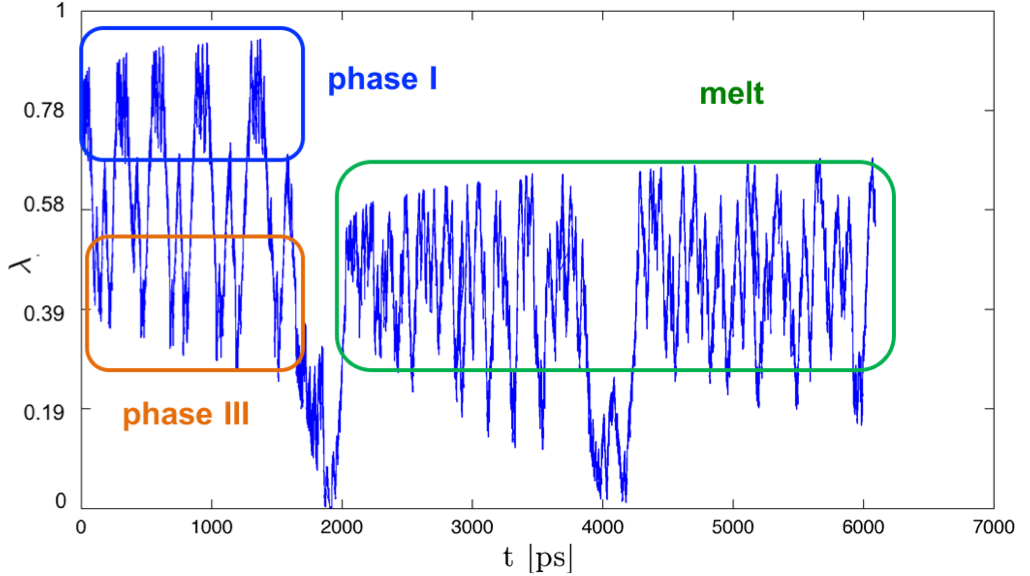}
\caption{Preliminary run at 600 K - 5 GPa considering $\lambda_I$ as the only CV. The results show that this parameter alone is unable to distinguish the basins of phase III and melt.}
\label{imageY2} 
\end{figure}

In the present work we employ a set of two CVs, namely $\lambda_I$ and $\lambda_{III}$. The necessity of two Cvs is due to the fact that in preliminary test runs with only $\lambda_I$, the basin of phase III overlaps with the one of the melt when it is formed close to the solid-liquid border (Figure~\ref{imageY2}).

\subsubsection{Unbiased MD}
\begin{figure}[t]
\includegraphics[width=0.51\textwidth]{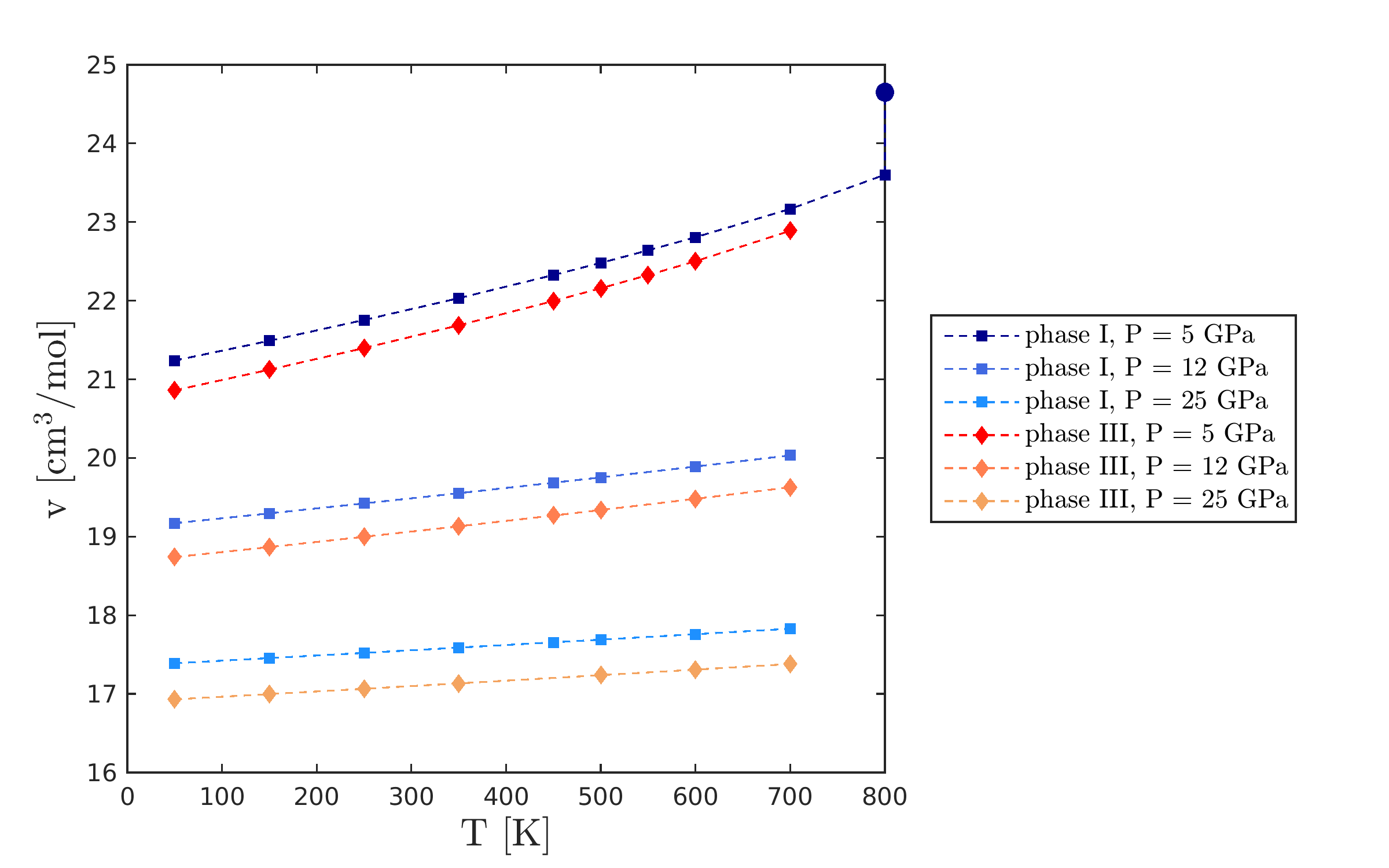}
\caption{Trend of the specific volume in cm$^3$/mol over temperature at different pressures (5, 12, 25 GPa) for phase I (squared markers) and phase III (diamond markers). The dark blue dot represents the value obtained for melt at 800 K - 5 GPa from the spontaneous evolution of phase I.}
\label{imagesTPV} 
\end{figure}

\begin{figure*}[t]
\includegraphics[width=0.7\textwidth]{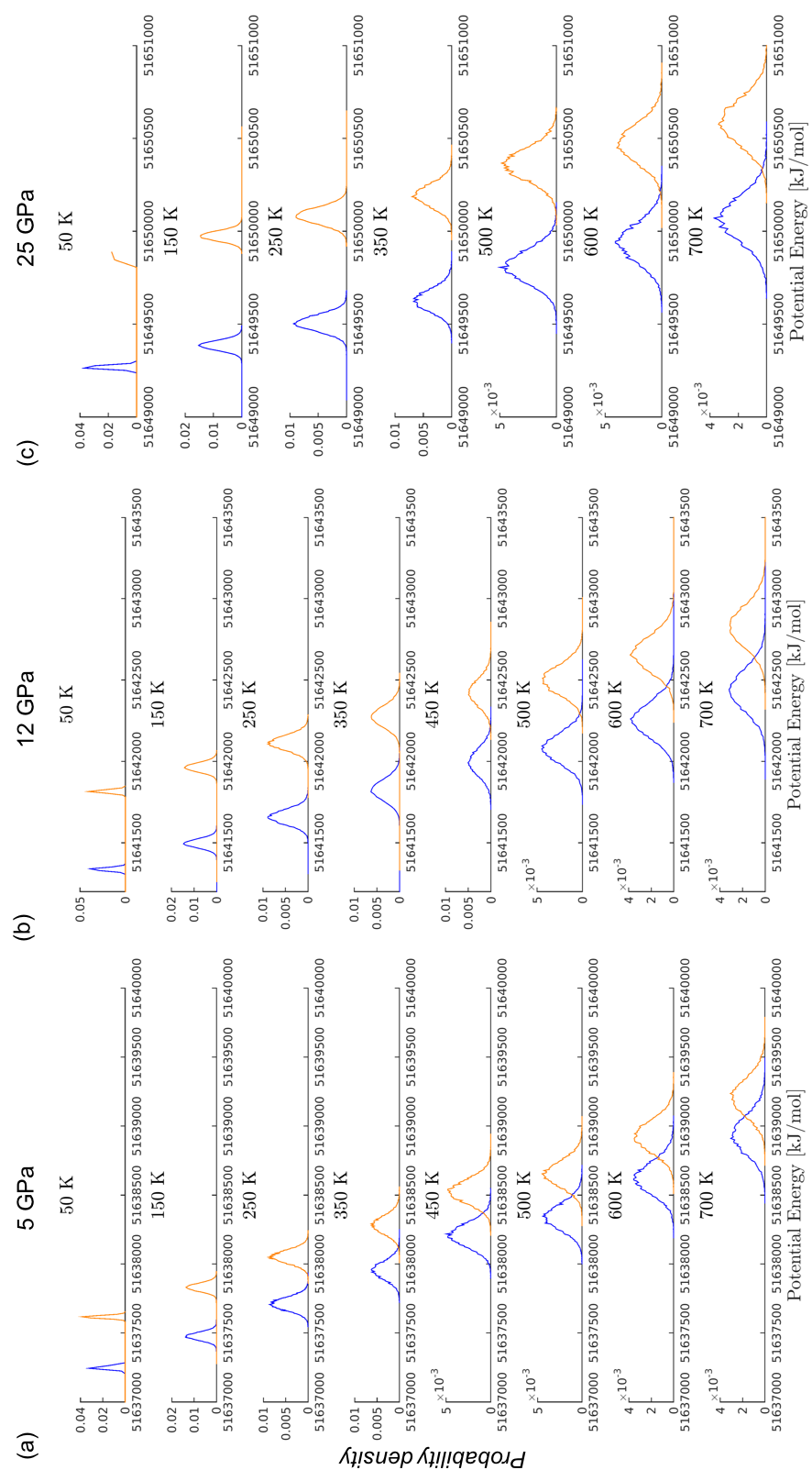}
\caption{Potential   energy    of    phase I (blue) and phase III (orange) in kJ/mol, expressed through   its   probability  distribution  over a range of temperature and pressure.}
\label{imagesTPE} 
\end{figure*}

\begin{figure*}[t]
\includegraphics[width=1\textwidth]{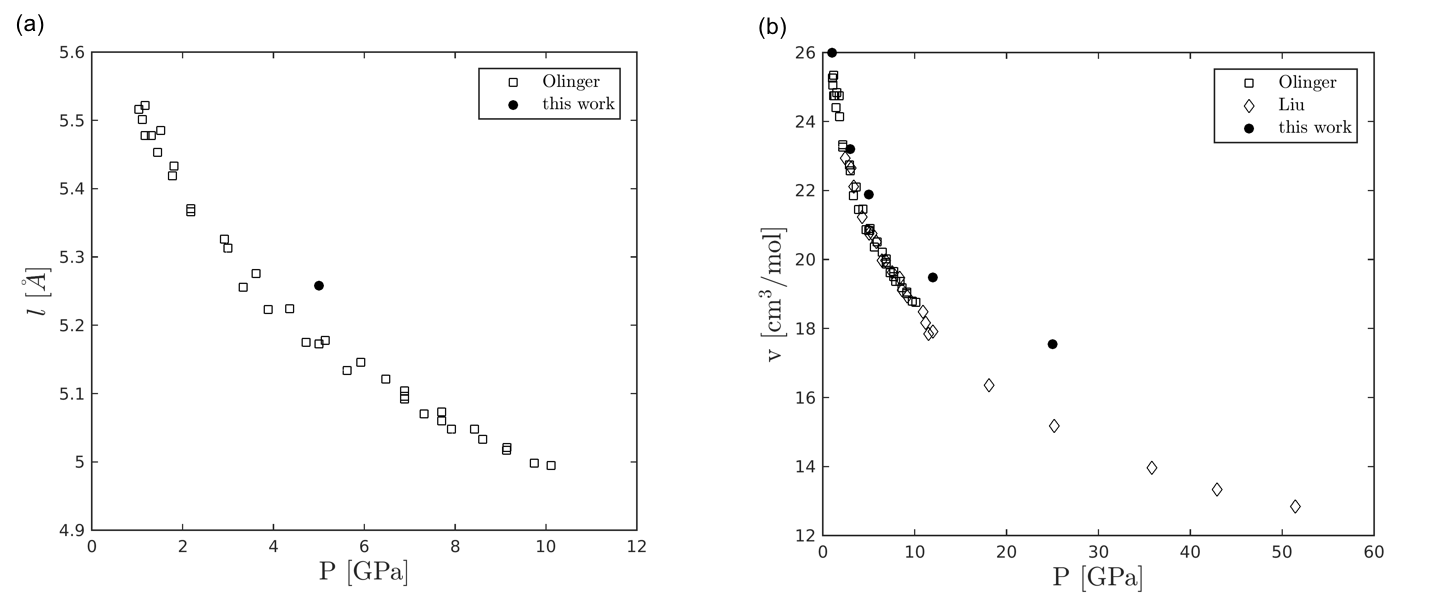}
\caption{Comparison of the unit cell side (a) and volume (b) of phase I with experimental results from Olinger\cite{Olinger1982}  and Liu\cite{Liu1984}.}
\label{imageComp} 
\end{figure*}

We hereafter report the results in terms of volume  (Figure~\ref{imagesTPV}) and potential energy (Figure~\ref{imagesTPE}) for unbiased MD  performed for phases I and III over a range of pressure (5 - 25 GPa) and temperature (50 - 700 K). The set up is the same described in the main text, a part from phase I, for which we employed a cut-­off  radius  of  1.0  nm. From these simulations we obtain the equilibrated initial configurations for WTMetaD as well as an insight into the properties of the system. \\
\newline The   general   trend   for volume and potential energy is   the   same   for both polymorphs:     volume   and   potential   energy   grow   with  temperature,   while   at   growing   pressures,   the   potential   energy   increases   and   the   volume   decreases.   Focusing on  the  volume (Figure~\ref{imagesTPV}),  orthorhombic  phase  III  has  a  smaller  volume  than cubic  I,  such difference being around  2  \%,  in  agreement  with  the  literature. MOreover, the comparison of the length of the unit cell edges of phase I with experimental results\cite{Olinger1982,Liu1984} (Figure~\ref{imageComp}) shows that the set up with the TraPPE force field tends to overestimate its value, with a smaller deviation the smaller the pressure. Our values at 295 K are interpolated from the reasonably linear trend shown by volume in Figure~\ref{imagesTPV}.
\newline The   width   of   the  distribution of potential energy, instead, increases with temperature and   such   enlargement   of   the   fluctuations   leads   to   an  interesting   overlap   of   the   potential   energy   distribution   of   the   two   polymorphs.   However, increasing   pressure   drastically   reduces   the   extent   of   the   overlay.   Moreover,   the   potential  energy  is  higher  for  phase  III  in  all  the  conditions  simulated.

In conclusion of the presentation of MD results, we highlight that no  transition  between  different  phases  is  sampled, unless in conditions of high overheating (phase I to melt at 800 K - 5 GPa, in agreement with  the   predictions   and   observations   of  P\'erez-S\'anchez et al.\cite{Perez-Sanchez2013}) or high under-pressurizing (phase III to I at 350 K - 1 GPa).  This  is  a  further  confirmation  that  polymorphic   transitions   are   rare   events,   which   require   enhanced   sampling   techniques   to   be  thoroughly  sampled.  In   addition,   we remark that, by   observing   the   potential   energy   distributions   of   phases   I   and   III,   such  property  by  itself  is  not  a  reliable  indicator  of  the  thermodynamic  stability.

\subsubsection{WTMetaD results}
\begin{figure*}[t]
\includegraphics[width=1\textwidth]{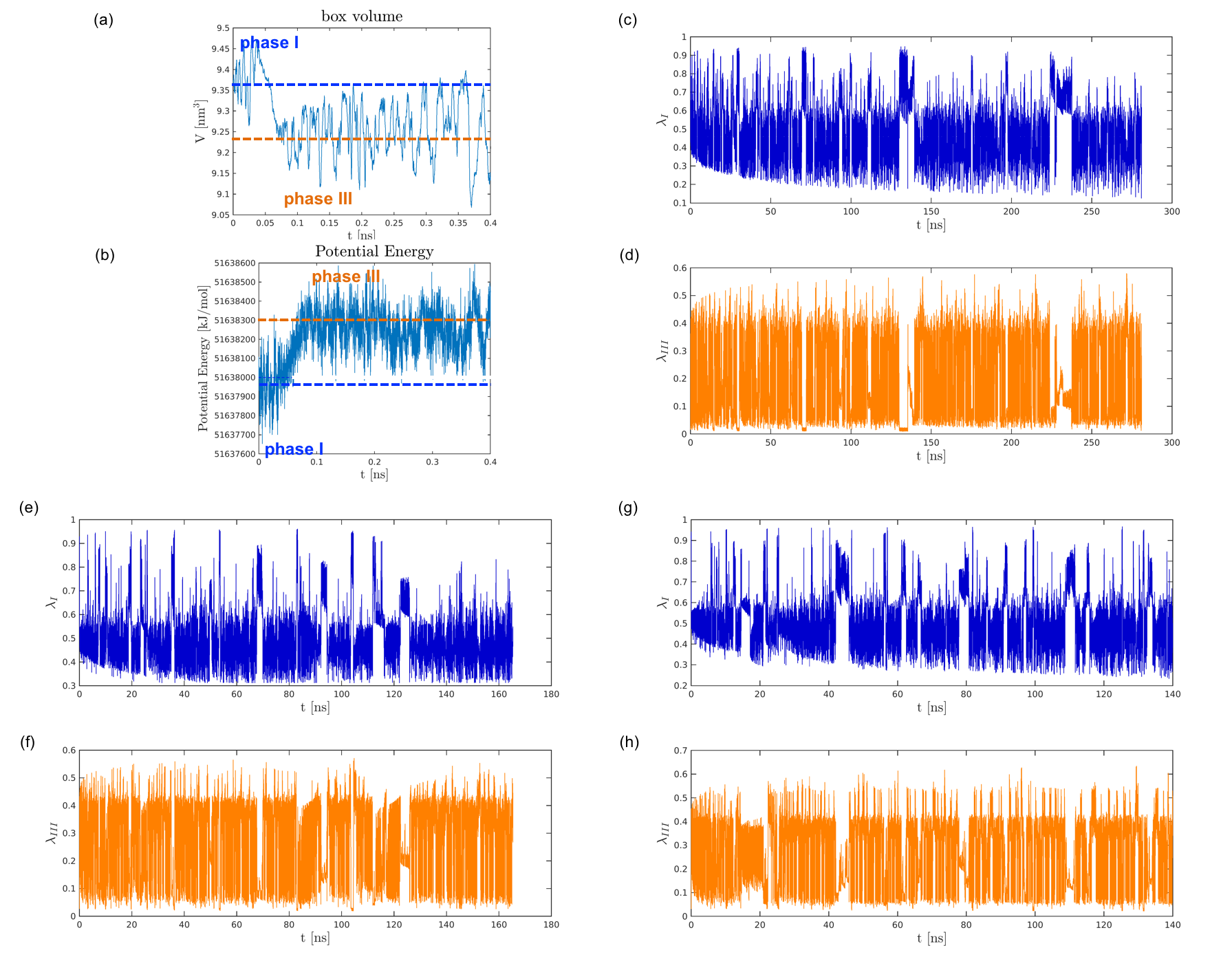}
\caption{Consistency check of the box volume (a) and the potential energy (b) obtain from WTMetaD with the unbiased values (dashed lines) of phases I (blue) and III (orange) for the explicative case at 350 K - 5 GPa; time evolution of the CVs at 350 K and 3 GPa (c-d), 8 GPa (e-f) and 12 GPa (g-h).}
\label{imageCons} 
\end{figure*}

In Figure~\ref{imageCons} (a-b), we report the fluctuations of the value of volume of the box and  potential energy during the first polymorphic transition at 350 K - 5 GPa, compared with the averages obtained in unbiased MD (dashed lines); as the results are in good agreement, WTMetaD gives an accurate description of the system.
\newline Then, Figure~\ref{imageCons} (c-h) plot the temporal evolution of $\lambda_I$ and $\lambda_{III}$ at 350 K and 3, 8, 12 GPa. Similar observations as for the case reported in the main text (Figure~\ref{imagetevolution}) can be drawn, in particular for the high number of conversions and the ``anisotropic'' transitions.


\subsubsection{Approximation of MFEP}
\begin{figure*}[t]
\includegraphics[width=1\textwidth]{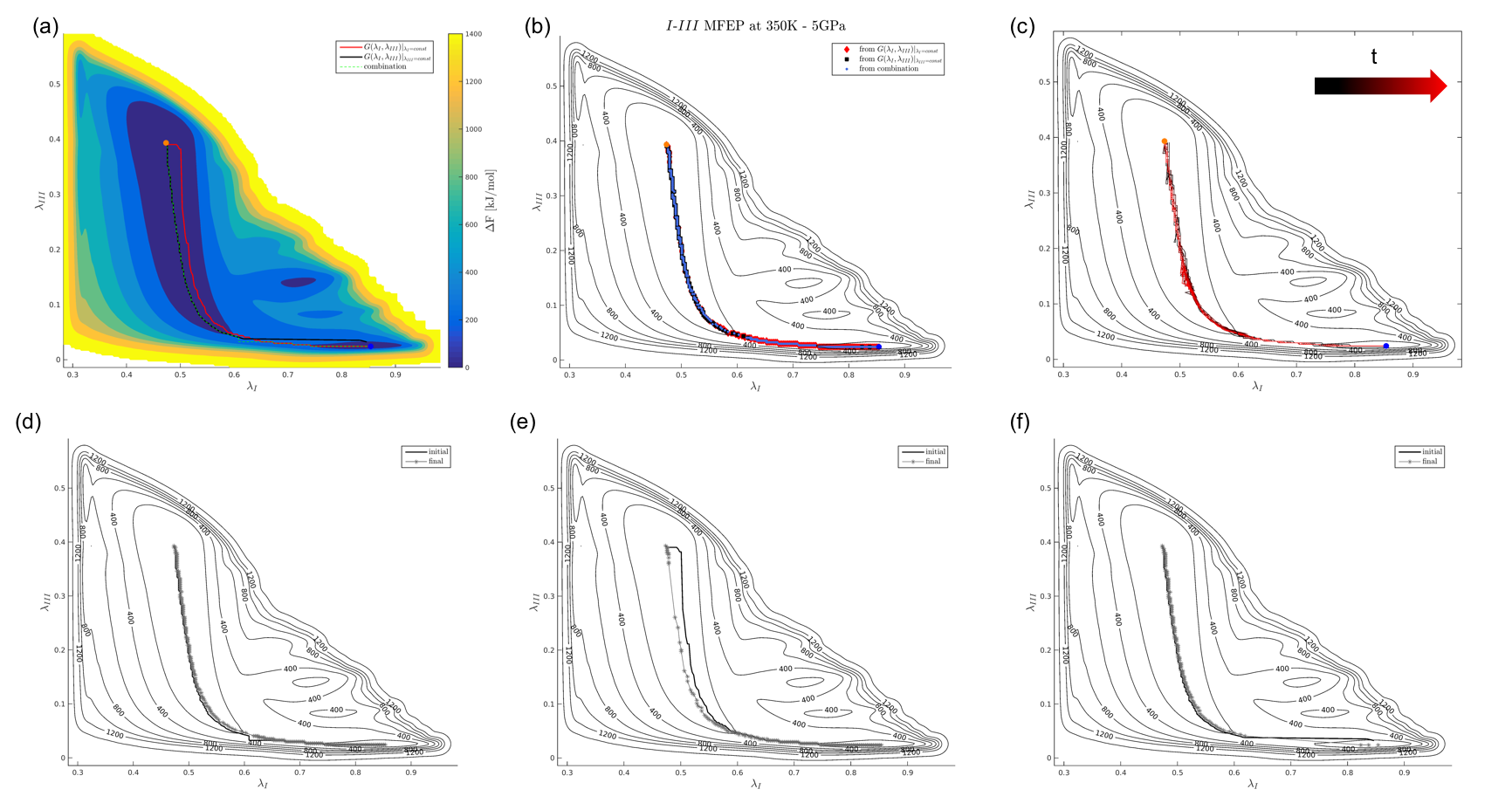}
\caption{(a) Comparison between transition pathways $G(\lambda_{I},\lambda_{III})\vert_{\lambda_I=const}$ (red line) and $G(\lambda_{I},\lambda_{III})\vert_{\lambda_{III}=const}$ (black line) and their combination (dashed green line) at 350 K - 5 GPa. (d)-(f) Initial guess pathway (black continuous line) and final MFEP (grey stars and line) obtained from our optimisation routine, starting from the combined path, $G(\lambda_{I},\lambda_{III})\vert_{\lambda_I=const}$ (red line) and $G(\lambda_{I},\lambda_{III})\vert_{\lambda_{III}=const}$, respectively. The comparison of the final results is reported in (b), which shows that the paths converge to the same actual MFEP. The time evolution of the MFEP is reported in (c), where the path colour changes from black to red with increasing simulation time.}
\label{imageMFEPappr} 
\end{figure*}

As educated guess of the MFEP we proposed a combination of two approximations of the transition pathway in CV space. Such approximations are namely $G(\lambda_{I},\lambda_{III})\vert_{\lambda_I=const}$, the locus of the minima in G($\lambda_I$,$\lambda_{III}$) at constant $\lambda_I$, and $G(\lambda_{I},\lambda_{III})\vert_{\lambda_{III}=const}$, the locus of the minima in G($\lambda_I$,$\lambda_{III}$) at constant $\lambda_{III}$. The necessity to combine these representations is due to the characteristic L-shaped FES, which causes $G(\lambda_{I},\lambda_{III})\vert_{\lambda_I=const}$ to be a better approximation of the MFEP within basin I, while $G(\lambda_{I},\lambda_{III})\vert_{\lambda_{III}=const}$ within basin III (Figure~\ref{imageMFEPappr} (a)).  
\newline We then performed an optimisation routine that, starting from an initial guess, attempts replacements of positions along the path with adjacent ones along specified direction: the trial move is accepted if the free energy is smaller than the initial guess and the coordinates are  different from the previous and following along the path; this procedure is iterated until convergence of the MFEP. The final path obtained is the same from different initial sets (Figure~\ref{imageMFEPappr}(b)). We notice that the MFEP is very similar to the combined path in the minima basins, while it describes more accurately the portion in between them (Figure~\ref{imageMFEPappr}(d)-(f)).
\newline The transition pathway converges to the MFEP well before the convergence of the simulation  (Figure~\ref{imageMFEPappr}(c)) and no alternative routes emerge.

\subsubsection{Committor Analysis}
\begin{figure*}[t]
\includegraphics[width=1\textwidth]{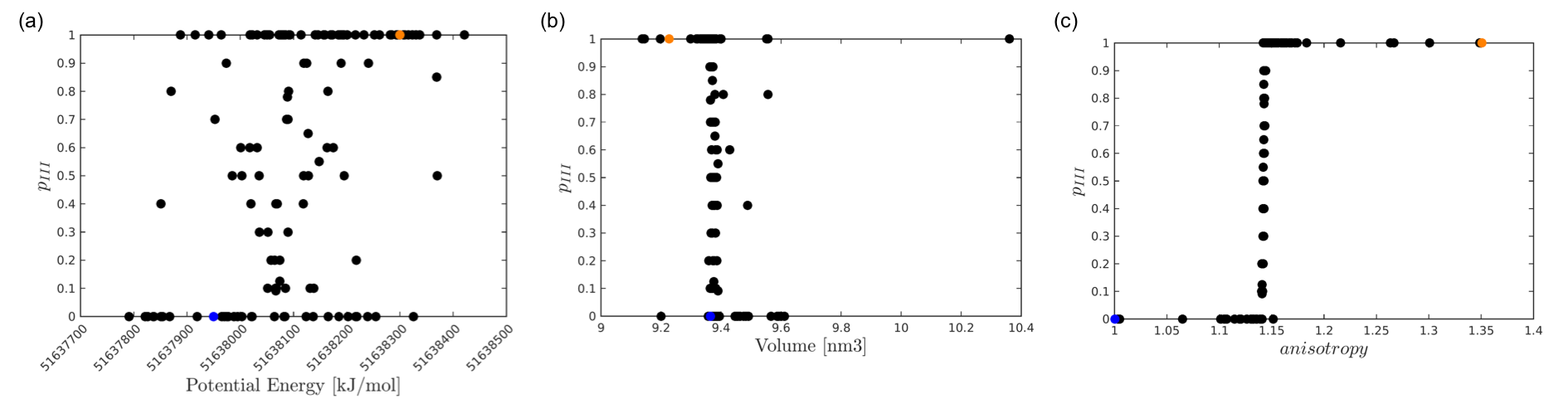}
\caption{Representation of the committor probability p$_{III}$ over potential energy (a), volume (b) and cell anisotropy (c)}
\label{imageCommAn} 
\end{figure*}

In order to better understand the area on the CV-space around the saddle point, we investigate the dependence of $p_{III}$ on properties other than the local order, in particular on potential energy, volume and anisotropy of the cell (Figure~\ref{imageCommAn}); the anisotropy is defined here as the ratio between the longest and shortest box edge. 
While the committor of a configuration does not seem to show any correlation with its potential energy or volume, the clear and steep sigma-shape that $p_{III}$ shows as a function of the box  anisotropy (Figure~\ref{imageCommAn} (c)) suggests the necessity to include this property in  the characterisation of the transition state.
